
\magnification=1200
\def\sect{\vskip 2mm \centerline}
\def\re{\hangindent=1pc  \noindent}
\def\cen{\centerline}
\def\v{\vskip 1mm}

\def\kms{km s$^{-1}$}
\def\deg{$^\circ$}
\def\Deg{^\circ}

\cen{\bf GIANT EXPLOSION AT THE GALACTIC CENTER AND}

\cen{\bf HUGE SHOCKED SHELLS IN THE HALO}

\v\v

\cen{Yoshiaki SOFUE}

\cen{\it Institute of Astronomy, University of Tokyo, Mitaka, Tokyo 181}

\cen {\it E-mail: sofue@sof.mtk.ioa.s.u-tokyo.ac.jp}

\cen{(To appear in Ap.J. Letter)}

\v\v
\cen{\bf Abstract}

We simulate the propagation of a shock front through the galactic halo induced
by an explosion and/or a starburst at the galactic center.
A huge dumbbell ($\Omega$)-shaped shock front  produced by an explosion of
energy $3\times 10^{56}$ ergs about fifteen million years ago can mimic the
radio and X-ray North Polar Spur as well the southern large X-ray spur.
The post-shock high-temperature gas in the corona will explain the observed
X-ray bulge around the galactic center.

Key workds Galaxy: center -- Galaxy: halo -- Radio continuum: general -- Shock
wave -- X-rays: general

\sect{I. GIANT EXPLOSION MODEL}\v

It is known that our Galactic Center has experienced various explosive events
(Oort 1977), and a number of expanding shells and rings have been observed,
associated with radio, X-ray and gaseous ejection of various scales (Sofue
1989).
In this letter we point out that the large-scale radio loops and X-ray diffuse
emission around the galactic center can be explained by a giant explosion
model.

The propagation of a shock front through the galactic halo induced by a point
energy injection at the center can be numerically simulated by applying the
self-similarity method (Sakashita 1971; Sofue 1984).
We assume an unperturbed density distribution of gas in the Galaxy as the
following:
$$\rho= \rho_0 e^{-\lbrace(\varpi/\varpi_0)^2+(z/z_0)^2 \rbrace} + \rho_{\rm
H}e^{-z/z_{\rm H}} + \rho_{\rm IG}, \eqno (1)$$
where $\varpi$ and $z$ are the distances from the rotation axis and the
galactic plane, respectively.
The first term represents the gas disk, and the constants are taken to be
$\rho_0=3$ H cm$^{-3}$,  $\varpi_0=7$ kpc, and $z_0=100$ pc.
The second term is a thick disk (or a halo) of atomic and molecular hydrogen,
which is indeed observed in several edge-on galaxies like NGC 891, and we take
$\rho_{\rm H}=0.03 $ H cm$^{-3}$ and  $z_{\rm H}=1$ kpc.
The third term is the intergalactic gas with a constant density of $\rho_{\rm
IG}=10^{-6}$ H cm$^{-3}$.
The formula gives a local hydrogen density of 1 H cm$^{-3}$ in the solar
vicinity.

The shock propagation has been calculated according to the method described by
Sofue (1984).
The result for an explosion energy of  $E=3\times 10^{56}$ is shown in  Fig. 1
and Fig. 2, where the tangential part of the shock front projected on the sky
is superposed on the observed X-ray maps and on the radio map.
The shock front at $b\sim 30\Deg$  reaches a radius of $ r = 5$ kpc at $t=1.5
\times 10^7$ years since the explosion, and the expansion velocity is $v \sim
200$ \kms.
 The calculated shock speed is higher in the polar direction than in the disk,
which yields the $\Omega$-shaped front.
The shock wave heats the halo gas up to $\sim 10^{6.5}$ K, and a
high-temperature corona remains in the post shock, which will be observed as an
X-ray bulge.
The emission measure toward the shock front at $b \sim 30\Deg$ is estimated to
be  0.03 cm$^{-6}$ pc.
The emission measure along the line tangential to the front is approximately
proportional to the mass plowed by the front from the galactic center,
therefore proportional to  cosec $b$, and is indeed observed in the M-band
intensity variation along the NPS (Fig. 3).
The cooling time of the hot gas is $ \sim 10^9$ years, and  so, the halo gas is
almost adiabatic.

\sect{II. RADIO AND X-RAY FEATURES}\v

The North Polar Spur (NPS) is a giant arc on the sky (Fig. 2), and has  been
studied on the assumption that it may be a local supernova remnant (SNR), that
had been proposed  three decades ago based on the earliest low-resolution data
(Berkhuijsen et al 1971; Spoelstra 1971; Egger 1993).
The SNR hypothesis has been criticized by  Sofue et al (1974), and some authors
have suggested that it could be a galactic scale phenomenon such as a galactic
helical magnetic field  (Mathewson et al 1968), or a magnetic inflation from
the galactic disk (Sofue 1973, 1976), and a giant shock  hypothesis has been
proposed by Sofue (1977, 1984).
We here revisit the NPS by applying an advanced data processing technique  to
the modern radio data (Haslam et al 1982), and combine them with the X-ray all
sky-survey.

Fig. 2 shows a new radio view of the NPS, which has been obtained by applying
the background-filtering technique (Sofue and Reich 1979) to the 408 MHz
whole-sky map (Haslam et al. 1982).
The NPS comprises a well-defined shock front at $l\sim 30\Deg$.
The radio ridge becomes sharper and narrower toward the galactic plane (Sofue
and Reich 1979), and the brightness  attains the maximum at $ ~b\sim 10\Deg$.
On this new radio view, the NPS and the major arcs along Loop IV (Berkhuijsen
et al 1971) can be naturally traced as one object, composing a giant Omega
anchored with its both roots to the galactic plane at $l\sim 20\Deg$ and $\sim
340\Deg$.
Negative-latitude spurs are seen at $l \sim 340\Deg$ to $ 320 \Deg,~ b \sim
-10\Deg$ to $-30\Deg$ and  at $l\sim 20\Deg, ~b\sim -2\Deg$ to $-40\Deg$,
though not clear.
We stress that the NPS can be approximately fitted by the calculated shock
front after $ t \sim 1.5\times10^7$.

Fig. 1 shows X-ray maps in the M (0.6 - 1.1 keV) and C (0.6 - 0.28 keV) bands
(McCammon et al 1983; McCammon and Sanders 1990).
The M-band map shows a global enhancement around the Galactic Center due to a
hot gas bulge (McCammon et al 1983), which indicates that the local HI disk is
transparent at $b> \sim 10\Deg$.
Enhancement along the NPS is evident in both bands.
A southern X-ray spur is also visible, particularly in the HEAO M-band map
(McCammon and Sanders 1990),  emerging from $(l,b) \sim (340\Deg,-10\Deg)$.
All these features appear to be well fitted by the calculated shock front.
The C-band map shows ``polar caps'' at high latitudes, which is also fitted by
the calculated shock front.
On the other hand, the C-band emission shows a wide absorption band along the
galactic plane, and the C-band NPS is also absorbed below $b=60\Deg$, hardly
visible below 30\deg.

Another remarkable feature in the M band is the sharp shadowing near the
galactic plane due to the local HI layer and the Hydra ridge (Cleary et al
1979).
Fig. 3 shows the variation of X-ray intensities along the NPS, where the
uniform background component has been subtracted and the values are normalized
to unitity at $b=60\Deg$.
Observed HI column density within $-70$ to +90 \kms\ is also plotted (Cleary et
al 1979) together with that for a plane-parallel model HI layer.
The M band emission increases toward the galactic plane, and appears to be
strongly absorbed at $b<10\Deg$.
This absorption feature shows an excellent correlation with the increasing HI
column density, and is naturally understood if the X-ray emitting region lies
beyond the HI disk further than 100 pc/sin 10\deg $\sim$ 600 pc.

Since the NPS at $b>60\Deg$ is clearly visible in the I (0.8 - 1.5 keV) band,
the temperature would be much higher than $10^6$ K.
Although a multi-temperature model may be required for a detailed fitting, we
here simply assume that the temperature is about 10$^{6.5}$ K.
Then, we can estimate the intrinsic (non-absorbed) C-band intensity relative to
the emission at $b= 60\Deg$, which  mimics the M-band distribution.
However, the observed C-band intensity is strongly absorbed below $b=60\Deg$
(Fig. 3), and the intensity at $b=30\Deg$ is only 0.09 times that of thus
expected intrinsic value.
Using the C-band transmission diagram (Fig. 11 of McCammon et al 1983), we can
estimate the corresponding HI mass to be  $7\times10^{20}$ H cm$^{-2}$, which
is greater than the observed value ($5\times10^{20}$ H cm$^{-3}$).
Hence, we may conclude that the X-rays at $b \sim 30\Deg$ originate {\it
beyond} the hydrogen layer.
As Fig. 3 indicates, the C-band emission is almost totally absorbed below
$b\sim 25\Deg$.
All these facts are consistent with the idea that the NPS lies in the halo
beyond the local disk.

According to the present model, the HI gas from the thick disk will be
accumulated in front of the shocked shell.
In the galactic plane it will make an expanding ring of a radius about 2.5 kpc,
and appears to coincide with the 3-kpc expanding ring (Oort 1977).
Above the galactic disk, an HI spur of intermediate velocity ($V_{\rm LSR}=30$
to 70 \kms) has been observed (Cleary et al 1979) at $l\sim 30\Deg, ~b\sim 10$
to 30\deg.
This velocity ($\sim 50$ \kms) corresponds to a kinematic distance of 4.4 kpc
and a galactocentric distance of 4.5 kpc.
Conservation of the angular momentum due to the large expansion of the shell
would result in a slower rotation at higher latitudes, where the accumulated HI
will be visible at lower or even negative LSR velocities.
In fact, a  low-velocity HI spur is observed  at $b>30\Deg$.
On the other hand, it seems difficult to explain the apparent alignment of
star-light polarization  along the HI ridge (Mathewson 1968; Mathewson and Ford
1970), if they are physically associated.
However, the physical alignment of star-light polarization along the
shock-compressed shell is controversial (Spoelstra 1971; Sofue et al. 1974).

\sect{III. DISCUSSION}\v

We have shown that many of the characteristics observed for the NPS and X-ray
features around the Galactic Center can be explained if the Galaxy has
experienced an active phase  $1.5 \times 10^7$ years ago associated with an
explosive energy release of some $10^{56}$ erg.
As to the origin of the energy, we may consider
(a) a giant explosion at the nucleus associated with the central massive black
hole; and/or
(b) a starburst which involved $\sim 10^5$ supernovae occurring in a relatively
short period, say, in $10^6$ yrs.
These two models are, however, essentially the same as far as the formation of
a shell of a large radius is concerned.
In our Galactic Center, various explosive events have been reported, and a
large number of expanding features are found, which could be the evidence for a
past starburst (Sofue 1989).
The total amount of energy is estimated to be about $10^{56-58}$ ergs (Oort
1977).
Explosion and starburst are not rare events in external galaxies:
radio shells of a few kpc scale in the halo are found in many spiral galaxies
such as  NGC 3079 (Duric et al 1983),  NGC 4258  (van Albada 1980), and NGC 253
 (Sofue 1984), which show an apparent similarity to the NPS.

The galactic-scale shock from the nuclear region will have an implication for
the evolution of the Galaxy:
The shock wave is an effective heating source of the hot corona.
It will contribute to a rapid circulation of heavy-elements from the inner disk
 to the entire Galaxy.
Finally we point out that the NPS as well as the X-ray spurs and the hot bulge
may become  tools to probe the galactic halo and its intergalactic interface.

\sect{References}\v

\re  Berkhuijsen, E., Haslam, C. G. T., and Salter, C. J. 1971, AA 14, 252.

\re  Cleary, M. N., Heiles, C., Haslam, C.G.T. 1979, AAS 36, 95.

\re  Duric, N.,  Seaquist, E.R.,  Crane, P.C.,  Bignell, R.C.,  Davis, L.E.,
1983,  Ap.J.L,   273,  L11.

\re Egger, R. 1993, Ph. D. Thesis, Univ. Munich, MPE Report No. 249.

\re  Haslam, C.G.T.,  Salter, C.J.,  Stoffel, H.,  Wilson, W.E.,  1982,  AA
Suppl. 47,  1.

\re  McCammon, D. and Sanders,  1990, ARAA 28, 657.

\re  McCammon, D., Burrows, D.N., Sanders.W. T., and Kraushaar, W. L. 1983,
Ap.J. 269, 107.

\re Mathewson, D. S. 1968, Ap.J. 153, L47.

\re Mathewson, D. S., Ford, V. L. 1970, Mem. RAS, 74, 139.

\re Sofue Y, K.Hamajima, Fujimoto, M 1974, PASJ, 26, 399

\re Sofue, Y. 1973, PASJ, 25, 207.

\re Sofue, Y. 1976, AA, 48, 1.

\re  Sofue Y 1977, AA 60, 327.

\re  Sofue, Y. 1984,  PASJ  36,  539.

\re  Sofue Y, Reich, W. 1979, AA Suppl. 38, 251.

\re  Sofue, Y. 1989,  in {\it The Center of the Galaxy} (ed. M. Morris, Kluwer
Academic Press, Dordrecht), p.213.

\re Spoelstra, T. A. T 1971, AA. 13, 273.

\re  van Albada, R. D. 1980, AA Suppl. 39, 283.

\v\v
\noindent {\bf Figure Captions}\v\v
\re Fig. 1: Calculated shock front in the galactic halo at 1, 1.5 and
$2\times10^7$ yr after an explosion and/or a starburst at the nucleus with a
total energy  of  $3 \times 10^{56}$ erg.
The front is superposed on the M- and C-band X-ray maps (McCammon 1983).

\v
\re Fig. 2a: The North Polar Spur at 408 MHz, showing a giant $\Omega$ over the
Galactic Cente, as obtained by subtracting the background from the Bonn-Parkes
all-sky survey (Haslam et al 1982) (top).
The grid interval is 20\deg in $(l, b)$ and (RA, Dec).
The calculated shock front is superimposed on the sketch of the NPS (bottom).

\v
\re Fig. 2b: The same as Fig. 2a but in a grey-scale representation.

 \v\re Fig. 3: C and M-band X-ray intensities along NPS normalized to unitity
at $b=60\Deg$ with the uniform background being subtracted.
The observed HI column density and that for a plane-parallel HI layer are
indicated.

\bye